\documentclass[12pt]{iopart}
\usepackage{iopams}  
\begin{document}

\title[Pressure formulas for liquid metals and plasmas]
{Pressure formulas for liquid metals and plasmas \\
based on the density-functional theory}

\author{Junzo Chihara\dag, 
Ichirou Fukumoto\ddag, Mitsuru Yamagiwa\ddag\  
and Hiroo Totsuji\S}

\address{\dag Research Organization for Information Science and Technology, 
Tokai, Ibaraki 319-1106, Japan\\
E-mail; chihara@ndc.tokai.jaeri.go.jp\\
\ddag Advanced Photon Research Center, JAERI,
Kizu, Kyoto 619-0215, Japan\\
\S Faculty of Engineering, Okayama University, Tsushimanaka 3-1-1, Okayama 700-8530, Japan}
\begin{abstract}
At first, pressure formulas for the electrons under the external potential produced 
by fixed nuclei are derived both in the surface integral and volume integral forms 
concerning an arbitrary volume chosen in the system; the surface integral form is described by 
a pressure tensor consisting of 
a sum of the kinetic and exchange-correlation parts in the density-functional theory, 
and the volume integral form represents 
the virial theorem with subtraction of the nuclear virial. Secondly 
on the basis of these formulas, the thermodynamical pressure of 
liquid metals and plasmas is represented in the forms of 
the surface integral and the volume integral including the nuclear contribution.
From these results, we obtain a virial pressure formula for liquid metals, which 
is more accurate and simpler than the standard representation.
From the view point of our formulation, some comments are made on 
pressure formulas derived previously and on a definition of pressure widely used.  
\end{abstract}




\section{Introduction}
In liquid metals and high density plasmas, their pressure 
is attributed in main to the behaviour of electrons involved in the system 
and the nuclear contribution is not so important. 
Therefore, almost pressure calculations take account of only 
the electron behaviour in the system; 
furthermore, the pressure is calculated as that of 
the electrons around one fixed nucleus neutralised in the volume (the Wigner-Seitz sphere). This model is called in several ways such as 
the ion-sphere model, the spherical shell model, the Wigner-Seitz model, or the confined atom model.  In the case of plasmas, the Thomas-Fermi approximation including its extention is applied for this purpose \cite{Feynman}-\cite{Ying89}, and in treating metals, the pressure is calculated by solving the wave equation based on the density-functional (DF) theory \cite{Liberman71}-\cite{LegPerrot01}.
As a consequence, the nuclear contribution to the pressure is not taken into account 
in almost cases. However, even in a high density plasma the nuclear contribution becomes important when its temperature is increased.
In the present work, we attempt to set up formulas to determine the pressure of liquid metals (plasmas) at arbitrary temperature and density by treating them as a nucleus-electron mixture using the virial and Hellmann-Feynman theorem.

There are many investigations 
\cite{Liberman71}-\cite{LegPerrot01} to set up 
the equation of states (EOS) from the nucleus-electron model 
on the basis of the virial theorem.
However, we find confusion and improper treatment in some works 
at the present stage. A typical and important example is the virial 
theorem for the electron pressure ${\bf P}_{\rm e}$ in the presence of fixed nuclei at the 
coordinate $\{{\bf R}_\alpha\}$ confined in a volume $\Omega$ \cite{Liberman71,Averill81}, 
which is described as
\begin{equation}\label{e:Janak}
3P_{\rm e}\Omega=2T_s[n]_\Omega+E_{\rm es}(\Omega)-\int_\Omega n({\bf r}){\bf r}
\nabla{\delta E_{\rm xc}\over\delta n({\bf r})} d{\bf r} \,.
\end{equation}
Here, $T_s[n]_\Omega$ is the kinetic energy defined in the DF theory 
for the system with a density distribution $n({\bf r})$;
$E_{\rm es}(\Omega)$ and $E_{\rm xc}[n]$ denote 
the electrostatic and exchange-correlation energies, respectively.
This equation should be written correctly, as will be discussed below, in the form 
\begin{eqnarray}\label{e:JanakC}
3P_{\rm e}\Omega&=&2T_s[n]_\Omega+E_{\rm es}(\Omega)
-\int_\Omega n({\bf r}){\bf r}\nabla{\delta E_{\rm xc}\over\delta n({\bf r})} d{\bf r}
\nonumber\\
&&-\sum_\alpha {\bf R}_{\alpha}\!\cdot\!{\bf F}_{\alpha}
+\oint_S{\bf r}\cdot{\bf P}_{\rm xc}^{\rm DF}\cdot d{\bf S}\,,
\end{eqnarray}
which reveals that there are no terms representing 
the nuclear virial $\sum_\alpha {\bf R}_{\alpha}\!\cdot\!{\bf F}_{\alpha}$ and a surface integral 
of the exchange-correlation pressure tensor ${\bf P}_{\rm xc}^{\rm DF}$ 
in the virial theorem given by (\ref{e:Janak}).  
The exact equation (\ref{e:JanakC}) involving these terms can be rewritten as
\begin{eqnarray}\label{e:JanakC2}
3P_{\rm e}\Omega
&=&2[T_s[n]_\Omega+\Delta T_{\rm xc}]+[E_{\rm es}(\Omega)
+\Delta U_{\rm xc}]-\sum_\alpha {\bf R}_{\alpha}\!\cdot\!{\bf F}_\alpha\\
&=&2T_\Omega+U_\Omega-\sum_\alpha {\bf R}_{\alpha}\!\cdot\!{\bf F}_\alpha\,, \label{e:PeC}
\end{eqnarray}
where $T_\Omega$ and $U_\Omega$ represent the true kinetic and potential 
energies, respectively, determined by treating the electrons as a many-electron system instead of the DF method. Here, the correction terms, 
$\Delta T_{\rm xc}$ and $\Delta U_{\rm xc}$, are related to the trace of the exchange-correlation pressure tensor ${\bf P}_{\rm xc}$ as follows
\begin{equation}
2\Delta T_{\rm xc}+\Delta U_{\rm xc}=\int_\Omega \tr {\bf P}_{\rm xc}d{\bf r}\,.
\end{equation}
On the other hand, the virial theorem (\ref{e:Janak}) 
due to the lack of the surface integral term
leads to the relation \cite{Averill81}
\begin{equation}
2\Delta T_{\rm xc}+\Delta U_{\rm xc}=-\int_\Omega n({\bf r}){\bf r}
\nabla{\delta E_{\rm xc}\over\delta n({\bf r})} d{\bf r}\,,
\end{equation}
which becomes zero for a uniform electron density inappropriately.

The nuclear virial term in (\ref{e:PeC}) causes some confusion in the calculation of EOS, 
since it has an inverse sign as to be considered 
as the nuclear pressure of the system, or it brings a difference 
from a naive virial theorem.
Following Slater \cite{SlaterBK}, many investigators including 
works of Janak \cite{Janak74}, 
Averill and Painter \cite{Averill81} and Ziesche, Grafenstein and 
Nielsen \cite{ZiescheGN88} have adopted a definition of the pressure 
in terms of the nuclear virial 
\begin{equation}\label{e:Pslater}
3P_{\rm e}\Omega\equiv \sum_\alpha {\bf R}_{\alpha}\!\cdot\!{\bf F}_\alpha
=2T_\Omega+U_\Omega\,.
\end{equation} 
The virial theorem (\ref{e:Janak}) is a result of this definition of the pressure. In the present work, we have proved 
that the electron pressure in the presence of nuclei fixed at $\{{\bf R}_\alpha\}$
should be given by (\ref{e:JanakC}) instead of (\ref{e:Pslater}).

On the other hand, another form of pressure definition is proposed 
by some authors \cite{NieMar85,Folland86,ZiescheGN88} with use of 
the Maxwell pressure tensor. However, this definition provides 
the real pressure only when some restriction is imposed, 
as will be discussed in the present work. In the next section, 
we pick up some important equations to discuss 
the virial theorem for a nucleus-electron mixture. Then, these equations are 
used to derive the electron pressure in the presence of nuclei 
in Sec.~\ref{s:EPressure}, and the thermodynamical pressure as a whole 
system in Sec.~\ref{s:ThP}. Some comments on other pressure formulas are 
made in Sec.~\ref{s:MaxP} and Sec.~\ref{s:comp}. The last section is devoted 
to summary and discussion.
 
\section{Exact relations to derive pressure formulas} 
For the purpose of preparing to derive pressure formulas, 
we describe in this section some 
necessary and important equations which can be proved without approximation.\\

\noindent (1) The DF theory:\\
Nuclei with the atomic number $Z$ fixed at positions $\{{\bf R}_\alpha\}$ causes an external potential $U_{\rm ext}({\bf r})$ for the electrons
\begin{equation}\label{e:Uext}
U_{\rm ext}({\bf r})\equiv -\sum_{\alpha=1}^N { Z e^2 \over |{\bf r}-
{\bf R}_{\alpha}| } ={\delta E_{\rm en}\over\delta n({\bf r})}\,. 
\end{equation}
Here, $E_{\rm en}$ is the electron-nucleus part of the total electrostatic energy $E_{\rm es}[n]$ for this system with an electron density distribution $n({\bf r})$
\begin{eqnarray}
E_{\rm es}[n]&\equiv &\frac{e^2}2\!\!\int {
 n({\bf r})n({\bf r'})\over |{\bf r}-{\bf r'}|}d{\bf r}d{\bf r'}
-\sum_{\alpha=1}^N\!\int {Z e^2 n({\bf r}')\over |{\bf r}'-
{\bf R}_{\alpha}| } d{\bf 
r}'+\frac12\sum_{\alpha\neq\beta}{Z^2 e^2 
\over |{\bf R}_{\alpha}-{\bf R}_{\beta}|} \label{e:EesA}\\
&\equiv &E_{\rm ee} +E_{\rm en} +E_{\rm nn}\,. \label{e:EesB}
\end{eqnarray}
The DF theory provides the effective external potential $U_{\rm eff}({\bf r})$ in terms of the intrinsic interaction part of the free-energy ${\cal F}_{\rm int}\equiv {\cal F}_{\rm xc}+E_{\rm ee} \label{e:Fxc}$ \cite{ChiharaDCF} 
\begin{equation}
U_{\rm eff}({\bf r})=U_{\rm ext}({\bf r})+{\delta {\cal F}_{\rm int}\over\delta
 n({\bf r})}-\mu_{\rm int}={\delta E_{\rm es}\over\delta n({\bf r})}
 +{\delta {\cal F}_{\rm xc}\over\delta n({\bf r})} -\mu_{\rm int}\,.
\label{e:Ueff}
\end{equation}
This effective external potential provides a way to calculate an exact electron density distribution $n({\bf r})$ 
by solving a wave equation for the one-electron wave function $\varphi_i({\bf r})$
\begin{equation}
\left[-\frac{\hbar^2}{2m}\nabla^2+U_{\rm eff}({\bf r}) 
\right]\varphi_i({\bf 
r})=\epsilon_i \varphi_i({\bf r})\,, \label{e:schDF}
\end{equation}
in the form
\begin{equation}
n({\bf r})=\sum_i f(\epsilon_i)|\varphi_i({\bf r})|^2\,, \label{e:dnsDF}
\end{equation}
with use of the Fermi distribution $f(\epsilon_i)$ for arbitrary temperature.\\

\noindent (2) Electrostatic energy $E_{\rm es}[n]$:\\
Since the interactions among electrons and nuclei are all Coulombic, 
the electrostatic energy $E_{\rm es}[n]$, (\ref{e:EesA}), is shown to 
satisfy the following equation
\begin{eqnarray}
E_{\rm es}&=&-\int n({\bf r}){\bf r}\nabla {\delta E_{\rm es}\over\delta n({\bf r})}d{\bf r}-\sum_\alpha {\bf R}_{\alpha}\nabla_\alpha E_{\rm es} \label{e:Ees1}\\
&=&-\int n({\bf r}){\bf r}\nabla U_{\rm eff}({\bf r})d{\bf r}
+\int n({\bf r}){\bf r}\nabla {\delta {\cal F}_{\rm xc}\over\delta n({\bf r})}d{\bf r}
-\sum_\alpha {\bf R}_{\alpha}\nabla_\alpha E_{\rm es}\,. \label{e:Ees2}
\end{eqnarray}
In the above, (\ref{e:Ees1}) is derived by noting the relation
\begin{equation}
{\bf r}\nabla v(|{\bf r}-{\bf r}'|)+{\bf r}'\nabla' v(|{\bf r}-{\bf r}'|)=|{\bf r}-{\bf r}'|v'(|{\bf r}-{\bf r}'|)\,,
\end{equation}
with $v'(r)\equiv \rmd v(r)/\rmd r$, and (\ref{e:Ees2}) is led by the use of (\ref{e:Ueff}). Equations, (\ref{e:Ees1}) and (\ref{e:Ees2}), are fundamental to derive the virial theorem for a nucleus-electron mixture.\\

\noindent (3) The kinetic energy in the DF theory:\\
In the DF theory, the kinetic energy $T_s[n]_\Omega$ of 
the electrons 
confined in an arbitrary finite volume $\Omega$ of the system is defined 
by the following equation (\ref{e:Ts1}) 
in terms of the wave functions for (\ref{e:schDF}), and it can be proved that 
this kinetic energy is associated with the external effective potential 
$U_{\rm eff}$ given by (\ref{e:Ueff}) and the density distribution $n({\bf r})$ under this 
external potential in the form
\begin{eqnarray}
2T_s[n]_\Omega&\equiv& 2\sum_i f(\epsilon_i)\int_\Omega 
\varphi^*_i({\bf r})\frac{-\hbar^2}{2m}\nabla^2
\varphi_i({\bf r})d{\bf r} \label{e:Ts1}\\
&=&\int_\Omega n({\bf r}){\bf r}\nabla U_{\rm eff}({\bf r})d{\bf r}
+\oint_S {\bf A} \cdot d{\bf S}\,. \label{e:Ts2} 
\end{eqnarray}
Here, the tensor ${\bf A}$ involved in the surface integral for the surface $S$ 
of the volume $\Omega$ is defined by the following equation (\ref{e:sur1}), 
which can be represented in other ways such as 
(\ref{e:sur2}) and (\ref{e:sur3}) in terms of the kinetic tensor 
$\tilde{\bf P}_{\rm K}^{\rm DF}$ defined by (\ref{e:sur3}) as follows
\begin{eqnarray}
\nabla \cdot {\bf A}&\equiv&\sum_i f(\epsilon_i)\frac{-\hbar^2}{2m}
\nabla\cdot\{\varphi_i^*\nabla[{\bf r}\nabla\varphi_i]
-({\bf r}\nabla\varphi_i)(\nabla\varphi_i^*)\} \label{e:sur1}\\
&=&\sum_i f(\epsilon_i)\frac{-\hbar^2}{4m}
\nabla\cdot\{\varphi_i^*\nabla[{\bf r}\nabla\varphi_i]
-({\bf r}\nabla\varphi_i)(\nabla\varphi_i^*)+{\rm c.c.}\} \label{e:sur2}\\
&=&\nabla\cdot [{\bf r}\cdot{\bf P}_{\rm K}^{\rm DF}]-\frac{\hbar^2}{4m}\nabla^2n({\bf r})
\equiv\nabla\cdot [{\bf r}\cdot\tilde{\bf P}_{\rm K}^{\rm DF}]\,, \label{e:sur3}
\end{eqnarray}
with
\begin{equation}
\fl {P}_{\rm K, \mu\nu}^{\rm DF}\equiv \sum_i f(\epsilon_i)\frac{-\hbar^2}{4m}
[\varphi_i^*\nabla_\mu\nabla_\nu\varphi_i+\varphi_i\nabla_\mu\nabla_\nu\varphi_i^*-(\nabla_\mu\varphi_i^*)(\nabla_\nu\varphi_i)
-(\nabla_\mu\varphi_i)(\nabla_\nu\varphi_i^*)]\,. \label{e:Pk}
\end{equation}
It should be noticed that these relations can be applied to 
a subspace $\Omega$ chosen arbitrarily in the whole space of the system.
The proof of the above relations is seen in Appendix.\\

\noindent (4) Pressure for a noninteracting electron gas in the external 
potential $U_{\rm eff}({\bf r})$:\\
The pressure tensor ${\bf P}_{\rm e}^0$ of a noninteracting electron gas in 
the external potential $U_{\rm eff}({\bf r})$ follows the next equation
\begin{equation}\label{e:Pe0Def}
\nabla \cdot {\bf P}_{\rm e}^0=-n({\bf r})\nabla U_{\rm eff}({\bf r})\,,
\end{equation}
which is a result of the local momentum balance. Here, note that $n({\bf r})$ 
is the density distribution of a noninteracting electron gas in the external 
potential $U_{\rm eff}({\bf r})$.\\

\noindent (5) Mathematical formula:\\
For arbitrary tensor ${\bf P}$, the following surface integral can be 
written in a volume integral form 
\begin{equation}
\int_\Omega \nabla \cdot ({\bf r}\cdot {\bf P})d{\bf r}=\oint_S
{\bf r}\cdot{\bf P}\cdot d{\bf S}=\int_\Omega \tr {\bf P}d{\bf r}
+\int_\Omega {\bf r}\cdot\nabla \cdot{\bf P}d{\bf r}\,. \label{e:tensor}
\end{equation}

\section{Electron pressure in the presence of fixed nuclei}\label{s:EPressure}
When the nuclei fixed at positions $\{{\bf R}_\alpha\}$ generate an inhomogeneous electron gas with the density distribution $n({\bf r})$,  
the electron pressure tensor ${\bf P}_{\rm e}$ of this system 
is defined by the following equation \cite{More79,Barto80}
\begin{equation}
\nabla\cdot {\bf P}_{\rm e}\equiv -n({\bf r})\nabla \!
\left[-\sum_{\alpha=1}^N\!{Z e^2\over |{\bf r}-
{\bf R}_{\alpha}| }+\int {
e^2 n({\bf r'})\over |{\bf r}-{\bf r'}|}d{\bf r'}\right]
=-n({\bf r})\nabla {\delta E_{\rm es}\over\delta n({\bf r})}\,.\label{e:PeDef}
\end{equation}
On the basis of this definition for ${\bf P}_{\rm e}$, we can prove 
the virial theorem for a subspace $\Omega$ 
arbitrarily chosen in the system as follows
\begin{eqnarray}
\oint_S{\bf r}\cdot{\bf P}_{\rm e}\cdot d{\bf S}&=&
\oint_S{\bf r}\cdot(\tilde{\bf P}_{\rm K}^{\rm DF}
+{\bf P}_{\rm xc}^{\rm DF})\cdot d{\bf S} \label{e:eeossur}\\
&=&2T_s[n]_\Omega+E_{\rm es}(\Omega)+\int_\Omega \tr {\bf P}_{\rm xc}d{\bf r}
-\sum_{\alpha\epsilon\Omega} {\bf R}_{\alpha}\!\cdot\!{\bf F}_\alpha \label{e:eeosvol}\\
&=&2T_\Omega+U_\Omega-\sum_{\alpha\epsilon\Omega} {\bf R}_{\alpha}\!\cdot\!{\bf F}_\alpha
=-\oint_S{\bf r}\cdot{\bf \sigma}_{\rm Bader}\cdot d{\bf S}\,. \label{e:eeosBad}
\end{eqnarray}
Here, ${\bf P}_{\rm xc}^{\rm DF}$ is the exchange-correlation 
pressure tensor defined in the DF theory as 
\begin{equation}\label{e:PxcDef}
\nabla\cdot {\bf P}_{\rm xc}^{\rm DF}\equiv n({\bf r})\nabla 
{\delta {\cal F}_{\rm xc}\over\delta n({\bf r})}\,.
\end{equation} 
which gives the explicit expression for the electrons in the jellium model at zero temperature in the local-density approximation (LDA) in the form 
\begin{equation}\label{e:PxcLDA}
{P}_{\rm xc,\mu\nu}^{\rm LDA}=n^2\frac{d\epsilon_{\rm xc}}{dn}\delta_{\mu\nu}\,,
\end{equation} 
with the exchange-correlation energy $\epsilon_{\rm xc}(n)$ per particle. 
Also, the force ${\bf F}_\alpha$ for $\alpha$-nucleus is introduced 
on the basis of the Hellmann-Feynman theorem: ${\bf F}_\alpha=-\nabla_{\alpha}E_{\rm es}$.
In (\ref{e:eeosBad}), the true kinetic and potential energies, 
$T_\Omega\equiv T_s+\Delta T_{\rm xc}$ and $U_\Omega
\equiv E_{\rm es}(\Omega)+\Delta U_{\rm xc}$, which should be provided by treating the many-electron problem, are obtained from the correction terms, $\Delta T_{\rm xc}$ and $\Delta U_{\rm xc}$; these correction terms are determined to satisfy the next relations
\begin{eqnarray}
2\Delta T_{\rm xc}+\Delta U_{\rm xc}&=&\int_\Omega \tr {\bf P}_{\rm xc}d{\bf r}\label{e:dPxc}\\
\Delta T_{\rm xc}+\Delta U_{\rm xc}&=&E_{\rm xc}\,.
\end{eqnarray}
In the LDA for the electrons in the jellium model, these correction terms are explicitly written as 
\begin{eqnarray}
\Delta T_{\rm xc}&=&\int_\Omega \tr {\bf P}_{\rm xc}d{\bf r}-E_{\rm xc} \approx \int_\Omega
 n({\bf r})\{3\mu_{\rm xc}({\bf r})-4\epsilon_{\rm xc}({\bf r})\}d{\bf r}\label{e:dTxc}\,,\\
\Delta U_{\rm xc}&=&2E_{\rm xc}-\int_\Omega \tr {\bf P}_{\rm xc}d{\bf r}\approx \int_\Omega n({\bf r})\{5\epsilon_{\rm xc}({\bf r})-3\mu_{\rm xc}({\bf r})\}d{\bf r}\, \label{e:dUxc}
\end{eqnarray}
with $\mu_{\rm xc}=\rmd(n\epsilon_{\rm xc})/\rmd n$. Furthermore, it is important to note that the virial theorem (\ref{e:eeosBad}) can be described in terms of the true kinetic stress tensor ${\bf \sigma}_{\rm Bader}$, which is defined by Bader \cite{SrebBader75} for the 
interacting electrons outside the framework of the DF theory; 
the proof of (\ref{e:eeosBad}) is given in \cite{SrebBader75}-\cite{Ghosh87}. Thus, the virial theorem with several expressions, (\ref{e:eeossur})-(\ref{e:eeosBad}), provides the electron pressure either in the form of surface integral (\ref{e:eeossur})
or volume integral (\ref{e:eeosvol}), if the volume is chosen so that the pressure is constant on the surface $S$ of $\Omega$.
At this point, it should be mentioned that the electron pressure in the fixed 
nuclei is represented by the volume integral form (\ref{e:eeosvol}) correctly if the virial of the forces which hold the nuclei is subtracted, while there is no need of this subtraction for the surface integral expression (\ref{e:eeossur}).

In the following, we give a brief outline of the proof of 
(\ref{e:eeossur})-(\ref{e:eeosBad}). At first, we can prove 
that the pressure tensor ${\bf P}_{\rm e}^0$, (\ref{e:Pe0Def}), of a 
noninteracting electron gas is identical with the kinetic pressure 
$\tilde{\bf P}_{\rm K}^{\rm DF}$, (\ref{e:sur3}), in the DF theory
\begin{equation}
{\bf P}_{\rm e}^0=\tilde{\bf P}_{\rm K}^{\rm DF}\,\label{e:PK0}
\end{equation}
with use of (\ref{e:tensor}), (\ref{e:Pe0Def}), (\ref{e:Ts2}) and (\ref{e:sur3}) as follows
\begin{eqnarray}
\oint_S{\bf r}\cdot{\bf P}_{\rm e}^0\cdot d{\bf S}
&=&\int_\Omega \tr {\bf P}_{\rm e}^0d{\bf r}
+\int_\Omega {\bf r}\cdot\nabla \cdot{\bf P}_{\rm e}^0d{\bf r}\label{e:PKdS0}\\
&=&2T_s[n]_\Omega-\int n({\bf r}){\bf r}\nabla U_{\rm eff}({\bf r})d{\bf r}\label{e:PKdS1}\\
&=&\oint_S{\bf r}\cdot\tilde{\bf P}_{\rm K}^{\rm DF}\cdot d{\bf S}\,.\label{e:PKdS}
\end{eqnarray}
Next, the electron pressure tensor ${\bf P}_{\rm e}$ is shown to be a sum of its kinetic pressure tensor $\tilde{\bf P}_{\rm K}^{\rm DF}$  and the exchange-correlation pressure tensor ${\bf P}_{\rm xc}^{\rm DF}$
\begin{equation}
{\bf P}_{\rm e}={\bf P}_{\rm e}^0+{\bf P}_{\rm xc}^{\rm DF}
=\tilde{\bf P}_{\rm K}^{\rm DF}+{\bf P}_{\rm xc}^{\rm DF}\,,\label{e:PeDF}
\end{equation}
from (\ref{e:Pe0Def}) and (\ref{e:PeDef}) as below
\begin{equation}
\fl \nabla \cdot ({\bf P}_{\rm e}-{\bf P}_{\rm e}^0)=n({\bf r})\nabla [ U_{\rm eff}({\bf r})-\delta E_{\rm es}/\delta n({\bf r})]=n({\bf r})\nabla 
[\delta {\cal F}_{\rm xc}/\delta n({\bf r})]
=\nabla\cdot {\bf P}_{\rm xc}^{\rm DF}\,.
\end{equation}
In final, with use of (\ref{e:PKdS1}), (\ref{e:PeDF}), (\ref{e:Ees2}) and (\ref{e:dPxc}), 
an identity (\ref{e:identity1}) is shown to arrive at the proof of (\ref{e:eeossur})-(\ref{e:eeosBad}) as is described in the following
\begin{eqnarray}
\hspace{-2cm}\oint_S{\bf r}\cdot{\bf P}_{\rm e}\cdot d{\bf S} 
&=&\oint_S{\bf r}\cdot{\bf P}_{\rm e}^0\cdot d{\bf S}
+\oint_S{\bf r}\cdot[{\bf P}_{\rm e}-{\bf P}_{\rm e}^0]\cdot d{\bf S}
=\oint_S{\bf r}\cdot(\tilde{\bf P}_{\rm K}^{\rm DF}
+{\bf P}_{\rm xc}^{\rm DF})\cdot d{\bf S} \label{e:identity1}\\
&=&2T_s[n]_\Omega-\int_\Omega n({\bf r}){\bf r}\nabla U_{\rm eff}({\bf r})d{\bf r}+\oint_S{\bf r}\cdot{\bf P}_{\rm xc}^{\rm DF}\cdot d{\bf S}\\
&=&2T_s[n]_\Omega+E_{\rm es}(\Omega)-\int_\Omega n({\bf r}){\bf r}\nabla{\delta {\cal F}_{\rm xc}\over\delta n({\bf r})} d{\bf r}
\nonumber\\
&&-\sum_{\alpha\epsilon\Omega} {\bf R}_{\alpha}\nabla_\alpha E_{\rm es}+\oint_S{\bf r}\cdot{\bf P}_{\rm xc}^{\rm DF}\cdot d{\bf S} \\
&=&2T_s[n]_\Omega+E_{\rm es}(\Omega)+\int_\Omega \tr {\bf P}_{\rm xc}d{\bf r}
-\sum_{\alpha\epsilon\Omega} {\bf R}_{\alpha}\!\cdot\!{\bf F}_\alpha \label{e:trPxc}\\
&=&2[T_s[n]_\Omega+\Delta T_{\rm xc}]+[E_{\rm es}(\Omega)+\Delta U_{\rm xc}]-\sum_{\alpha\epsilon\Omega} {\bf R}_{\alpha}\!\cdot\!{\bf F}_\alpha\,.
\end{eqnarray} 
In the derivation of (\ref{e:trPxc}), Eq.~(\ref{e:PxcDef}) and the tensor relation (\ref{e:tensor}) are used in the form
\begin{equation} 
\fl -\int n({\bf r}){\bf r}\nabla{\delta {\cal F}_{\rm xc}\over\delta n({\bf r})} d{\bf r}+\oint_S{\bf r}\cdot{\bf P}_{\rm xc}^{\rm DF}\cdot d{\bf S}=-\int_\Omega {\bf r}\cdot\nabla \cdot{\bf P}_{\rm xc}^{\rm DF}d{\bf r}+\oint_S{\bf r}\cdot{\bf P}_{\rm xc}^{\rm DF}\cdot d{\bf S}=\int_\Omega \tr {\bf P}_{\rm xc}^{\rm DF}d{\bf r}\,,
\end{equation}
which is written in the LDA for the electrons in the jellium model as
\begin{equation}
\fl -\int n({\bf r}){\bf r}\nabla{\delta {\cal F}_{\rm xc}\over\delta n({\bf r})} d{\bf r}=-\int n({\bf r}){\bf r}\nabla\mu_{\rm xc}d{\bf r}
=3\int n^2\frac{d}{dn}\epsilon_{\rm xc}d{\bf r}
-\oint n^2\frac{d}{dn}\epsilon_{\rm xc}{\bf r}\cdot d{\bf S}\,.
\end{equation}

\section{Maxwell stress tensor and electron pressure tensor}\label{s:MaxP}
In the previous section, we have shown that the electron pressure tensor 
is given by ${\bf P}_e\equiv \tilde{\bf P}_{\rm K}^{\rm DF}+{\bf P}_{\rm xc}^{\rm DF}$ in the DF theory. On the other hand,
Nielsen and Martin \cite{NieMar85} and some other \cite{Folland86,ZiescheGN88} have defined the electron pressure tensor ${\bf P}_e^{\rm NM}$ using the 
Maxwell pressure tensor as to be 
\begin{equation}\label{e:PeNM0}
{\bf P}_e^{\rm NM}\equiv \tilde{\bf P}_{\rm K}^{\rm DF}+{\bf P}_{\rm xc}^{\rm DF}+{\bf P}^{\rm M}\,.
\end{equation}
In this section, we inquire into the relationship between these two pressure tensors.

When the nuclei is fixed at the positions $\{{\bf R}_\alpha\}$ 
with a charge density distribution $\rho_+({\bf r})\equiv \sum_\alpha 
Ze\delta ({\bf r}-{\bf R}_\alpha)$, 
and this nuclear distribution produces an electron charge 
distribution $\rho_{-}({\bf r})\equiv -en({\bf r})$, 
an electrostatic potential $\phi({\bf r})$ of this system is determined by the Poisson equation
\begin{equation}\label{e:poisson} 
\nabla^2 \phi({\bf r})=-4\pi [\rho_{+}({\bf r})+\rho_{-}({\bf r})]\,.
\end{equation}
In terms of the electric field ${\bf E}({\bf r})=-\nabla \phi({\bf r})$ determined by this Poisson equation, we can define the Maxwell stress tensor ${\bf T}$ as 
\begin{equation}
{\bf T}\equiv \frac1{4\pi}\left(\nabla\phi\cdot\nabla\phi-\frac12|\nabla\phi|^2{\bf 1}\right)=\frac1{4\pi}\left({\bf E}\cdot{\bf E}-\frac12|{\bf E}|^2{\bf 1}\right)\,.
\end{equation}
Here, $\phi({\bf r})$ is represented as a solution of (\ref{e:poisson}) in the form
\begin{equation}\label{e:potP1}
\fl \phi({\bf r})=\!\int {[\rho_{+}({\bf r}')+\rho_{-}({\bf r}')]
 \over |{\bf r}-{\bf r'}|}d{\bf r'}=\sum_{\alpha=1}^N\!{Z e\over |{\bf r}-
{\bf R}_{\alpha}| }-\int {
e n({\bf r'})\over |{\bf r}-{\bf r'}|}d{\bf r'}\equiv 
{\Phi ({\bf r})\over -e}\,
\end{equation}
with
\begin{equation}\label{e:potP2}
\Phi ({\bf r})\equiv \int {
e^2 n({\bf r'})\over |{\bf r}-{\bf r'}|}d{\bf r'}
-\sum_{\alpha=1}^N\!{Z e^2\over |{\bf r}-
{\bf R}_{\alpha}| }={\delta E_{\rm es}\over\delta n({\bf r})}\,.
\end{equation}
The Maxwell stress tensor is shown to satisfy the following equation
\begin{equation} \label{e:DivMax}
\nabla\cdot{\bf T}=-[\rho_{+}({\bf r})+\rho_{-}({\bf r})]\nabla \phi({\bf r})
=\frac1{e}[\rho_{+}({\bf r})+\rho_{-}({\bf r})]\nabla \Phi({\bf r})\,.\end{equation}
With use of (\ref{e:potP2}), (\ref{e:DivMax}) and 
(\ref{e:tensor}), we obtain the relations between the Maxwell stress tensor ${\bf T}$ and the electrostatic energy $E_{\rm es}$, (\ref{e:EesA}), of the system (excluding nuclear self-interactions)
\begin{eqnarray}
\int_\Omega {\bf r}\cdot\nabla \cdot{\bf T}d{\bf r}&=&-\int_\Omega n({\bf r}){\bf r}\nabla {\delta E_{\rm es}\over\delta n({\bf r})}d{\bf r}-\sum_{\alpha\epsilon\Omega} {\bf R}_{\alpha}\nabla_\alpha E_{\rm es} \label{e:EesT1}\\
&=&E_{\rm es}(\Omega)=-\int_\Omega \tr {\bf T}d{\bf r}+\oint_S
{\bf r}\cdot{\bf T}\cdot d{\bf S}\,. \label{e:EesT2}
\end{eqnarray}
Thus, we obtain the following relation which is important to know how the 
Maxwell stress tensor ${\bf T}$ plays a role in the virial theorem  
\begin{eqnarray}
\fl -\!\int_\Omega\!\tr {\bf T}d{\bf r}&=&\frac1{8\pi}\int_\Omega |{\bf E}|^2d{\bf r}
=-\int_\Omega {
 [\rho_+({\bf r})\!+\!\rho_-({\bf r})]\ {\bf r}\cdot\nabla\phi({\bf r})}d{\bf r}d{\bf r'}-\oint_S
{\bf r}\cdot{\bf T}\cdot d{\bf S}\\
\hspace{-2.5cm}&=&E_{\rm es}(\Omega)-\oint_S
{\bf r}\cdot{\bf T}\cdot d{\bf S}\,. \label{e:trE}
\end{eqnarray}

At this stage, we introduce the Maxwell pressure tensor 
${\bf P}^{\rm M}\equiv -{\bf T}$; then, we obtain the following relations 
from (\ref{e:DivMax}), (\ref{e:potP2}) and (\ref{e:PeDef})
\begin{equation}
-\nabla \cdot (\tilde{\bf P}_{\rm K}^{\rm DF}+{\bf P}_{\rm xc}^{\rm DF}+{\bf P}^{\rm M})=\sum_\alpha{\bf F}_\alpha\delta ({\bf r}-{\bf R}_\alpha)\,, \label{e:MaxF}
\end{equation}
\begin{equation}
-\int_\Omega {\bf r}\cdot\nabla \cdot (\tilde{\bf P}_{\rm K}^{\rm DF}+
{\bf P}_{\rm xc}^{\rm DF}+{\bf P}^{\rm M})d{\bf r}
=\sum_{\alpha\epsilon\Omega}{\bf R}_\alpha\cdot{\bf F}_\alpha\,.
\end{equation}
Therefore, we get an expression with use of (\ref{e:trE})
\begin{equation}
\int_\Omega \tr (\tilde{\bf P}_{\rm K}^{\rm DF}
+{\bf P}_{\rm xc}^{\rm DF}+{\bf P}^{\rm M})d{\bf r}
=2T_\Omega+U_\Omega
+\oint_S
{\bf r}\cdot{\bf P}^{\rm M}\cdot d{\bf S}\,, \label{e:virMAX}
\end{equation}
for an arbitrary subspace $\Omega$ chosen in the system.
It should be mentioned that in the standard treatment the surface 
integral in (\ref{e:trE}) is neglected, and as a result, the surface 
integral of the Maxwell pressure tensor does not appear in (\ref{e:virMAX}); 
Eq.~(\ref{e:virMAX}) without the surface integral term implies that
${\bf P}_e^{\rm NM}$ defined by (\ref{e:PeNM0}) seems to be interpreted as the electron 
pressure tensor giving the virial theorem exactly \cite{Folland86,ZiescheGN88}. This is not the case, since the surface integral exists in (\ref{e:virMAX}) in general. At the same time, 
it should be kept in mind that we have the virial theorem for ${\bf P}_e^{\rm NM}$
\begin{equation}\label{e:virMAXs}
\oint_S {\bf r}\cdot (\tilde{\bf P}_{\rm K}^{\rm DF}+
{\bf P}_{\rm xc}^{\rm DF}+{\bf P}^{\rm M})\cdot d{\bf S}
=2T_\Omega+U_\Omega-\sum_{\alpha\epsilon\Omega}{\bf R}_{\alpha}\!\cdot\!{\bf F}_\alpha\,.
\end{equation}
only if a volume $\Omega$ is chosen so as to fulfill 
the condition $\oint_S
{\bf r}\cdot{\bf P}^{\rm M}\cdot d{\bf S}=0$ on the surface of this volume; 
in this case, (\ref{e:virMAXs}) reduces to (\ref{e:eeossur}).

In conclusion, as a consequence of (\ref{e:virMAX}) and (\ref{e:virMAXs}), the tensor 
${\bf P}_e^{\rm NM}$ defined by (\ref{e:PeNM0}) plays a role of 
the electron pressure tensor 
in the presence of the fixed nuclei only for a subspace $\Omega$, on which surface 
the condition $\oint_S
{\bf r}\cdot{\bf P}^{\rm M}\cdot d{\bf S}=0$ is satisfied.

\section{Comparison with other electron-pressure formulas}\label{s:comp}
In Sec. \ref{s:EPressure}, we derived the virial theorem 
(\ref{e:eeossur})-(\ref{e:eeosBad}), in the surface integral and volume 
integral forms, for the electrons in the presence of fixed nuclei, 
which provides the electron pressure in condensed matter at 
arbitrary temperature and density. On the other hand, there have been proposed 
many methods to calculate the electron-pressure in the fixed nuclei. 
In the present section, we will make comparison of our result with some of 
other formulations, and give some comments.

If the electron pressure is assumed to be constant on the surface of 
the Wigner-Seitz sphere, Eq.~(\ref{e:eeossur}) with (\ref{e:sur2}) 
provides the electron pressure 
in the presence of the fixed nuclei in a surface integral form
\begin{eqnarray}\label{e:Liber1}
\fl \oint_S{\bf r}\cdot{\bf P}_{\rm e}\cdot d{\bf S}&=&3P_{\rm e}V
=\oint_S{\bf r}\cdot(\tilde{\bf P}_{\rm K}^{\rm DF}
+{\bf P}_{\rm xc}^{\rm DF})\cdot d{\bf S}\nonumber\\
&=&\sum_i f(\epsilon_i)\frac{-\hbar^2}{4m}
\oint_S\{\varphi_i^*\nabla[{\bf r}\nabla\varphi_i]
-({\bf r}\nabla\varphi_i)(\nabla\varphi_i^*)+{\rm c.c.}\}dS\nonumber\\
&&+\oint_S{\bf r}\cdot{\bf P}_{\rm xc}^{\rm DF}\cdot d{\bf S}\,.
\end{eqnarray}
When we use  the LDA expression for $P_{\rm xc,\mu\nu}^{\rm DF} \simeq n^2\rmd \epsilon_{\rm xc}/\rmd n\delta_{\mu\nu}$ in the above, (\ref{e:Liber1}) reduces to Liberman's pressure formula. 
However, the nuclear virial term does not appear in (\ref{e:Liber1}) 
on comparison with the Liberman expression (7).

On the other hand, if we choose the volume $\Omega$ to be the 
Wigner-Seitz sphere with a radius $r_0$, 
the virial theorem in the surface integral form, (\ref{e:eeossur}), with (\ref{e:Pk}) can be rewritten as
\begin{eqnarray}\label{e:Pett}
\hspace{-1.5cm}P_{\rm e}&=&\sum_i f(\epsilon_i)
\oint_S\varphi_i^*(r_0,\hat {\bf r})\left[\epsilon_i
-U_{\rm eff}-\frac{\hat \ell^2}{2mr_0^2} \right]\varphi_i(r_0,\hat {\bf r})\frac{d\Omega}{4\pi}\nonumber\\
\hspace{-1.5cm}&&+\sum_i f(\epsilon_i)\frac{\hbar^2}{2m}
\oint_S\left|\frac{\varphi_i}{r_0}\right|^2\left[ \left|r\frac{\partial\varphi_i}{\partial r}/\varphi_i\right|^2-\Re\left\{r\frac{\partial\varphi_i}{\partial r}/\varphi_i\right\}\right]_{r_0}\frac{d\Omega}{4\pi}
+P_{\rm xc}^{\rm DF}(r_0)
\end{eqnarray}
with $d\Omega\equiv\sin\theta d\theta\phi$ and the definition of 
an operator $\hat\ell^2$: $\hat\ell^2Y_\ell^m(\theta,\phi)=\ell(\ell+1)\hbar^2Y_\ell^m(\theta,\phi)$.\\
Equation (\ref{e:Pett}) becomes the pressure formula given by Nieminen-Hodge \cite{NiemHodg76} and Pettifor \cite{Pettifor76}. Furthermore, 
when the volume is chosen to be the Wigner-Seitz cell with the muffin-tin  approximation, and the exchange-correlation effect is described by a generalized gradient approximation, Eq.~(\ref{e:Pett}) leads to the  pressure formula given by Legrand and Perrot \cite{LegPerrot01} by taking account of the discontinuity in the electron density distribution.

Nielsen and Martin \cite{NieMar85} with other investigators \cite{Folland86,ZiescheGN88} have defined the pressure in terms of the Maxwell pressure tensor in the form
\begin{equation}
{\bf P}_e^{\rm NM}\equiv \tilde{\bf P}_{\rm K}^{\rm DF}+{\bf P}_{\rm xc}^{\rm DF}+{\bf P}^{\rm M}\,.
\end{equation}
As was discussed in details in Sec.~\ref{s:MaxP}, this tensor 
${\bf P}_e^{\rm NM}$ cannot be interpreted as the electron pressure 
except the case when we treat the pressure in a restricted subspace 
$\Omega$ chosen to satisfy the condition 
$\oint_S {\bf r}\cdot{\bf P}^{\rm M}\cdot d{\bf S}=0$ on its surface (e.g. Wigner-Seitz cell). When this condition is fulfilled, we obtain virial expressions such as 
\begin{eqnarray}
\int_\Omega {\rm tr} {\bf P}_e^{\rm NM}d{\bf r}&=&2T_\Omega+U_\Omega\,,\\
\oint_S {\bf r}\cdot {\bf P}_e^{\rm NM}\cdot d{\bf S}&=&2T_\Omega+U_\Omega-\sum_{\alpha\epsilon\Omega} {\bf R}_{\alpha}\!\cdot\!{\bf F}_\alpha\,.
\end{eqnarray}
In addition, Ziesche, Gr\"afenstein and Nielsen \cite{ZiescheGN88} have defined the pressure in terms of 
$\sum_\alpha {\bf R}_{\alpha}\!\cdot\!{\bf F}_\alpha$ considering a solid 
as an isolated system [that is, 
$\oint_S {\bf r}\cdot {\bf P}^{\rm NM}\cdot d{\bf S}=0$] by 
\begin{equation}
3P_{\rm e}\Omega\equiv\sum_\alpha {\bf R}_{\alpha}\!\cdot\!{\bf F}_\alpha
=2T_\Omega+U_\Omega\,,
\end{equation}
as was defined by Slater \cite{SlaterBK}.
However, the electron pressure in the fixed nuclei should be defined by 
(\ref{e:eeosBad}), where the virial of the forces that hold the nuclei fixed 
is subtracted here to be added later
in the thermodynamic pressure formula (\ref{e:truevir0}) as will be discussed 
in the next section.

Janak \cite{Janak74} derived the pressure formula starting from the 
virial theorem of the volume integral form which is written in our version as
\begin{equation}\label{e:virvolJanak}
3PV=2T_s[n]+E_{\rm es}+\int_V {\rm tr}{\bf P}_{\rm xc}d{\bf r}\,.
\end{equation}
In this equation, he replaced $E_{\rm es}$ in a volume $V$ containing $N$ nuclei by an approximate expression (\ref{e:EesApp}) 
\begin{eqnarray}
\fl Nu_{\rm es}&\equiv&E_{\rm es}\\
\fl &=&-\int_V n({\bf r}){\bf r}\nabla U_{\rm eff}({\bf r})d{\bf r}
+\int_V n({\bf r}){\bf r}\nabla {\delta E_{\rm xc}\over\delta n({\bf r})}d{\bf r}
-\sum_\alpha {\bf R}_{\alpha}\nabla_\alpha E_{\rm es}\\
\fl &\simeq& N\left[-\int_{\Omega_0} n({\bf r}){\bf r}\nabla U_{\rm eff}({\bf r})d{\bf r}
+\int_{\Omega_0} n({\bf r}){\bf r}\nabla {\delta E_{\rm xc}\over\delta n({\bf r})}d{\bf r}
-\frac14C\frac{Z_{\rm out}^2}{a}\right]\,, \label{e:EesApp}
\end{eqnarray}
as was written in his equation (18) for the unit cell $\Omega_0$.
Thus, the Ewald term $-\frac14C\frac{Z_{\rm out}^2}{a}$ can be taken as the nuclear virial in this approximation.
With this replacement, Eq.~(\ref{e:virvolJanak}) 
is rewritten as the next equation 
(\ref{e:eossurJ}), which leads to a simplified expression of 
his pressure formula [his (24)] as (\ref{e:PJanak}) 
\begin{eqnarray} 
3PV
&=&\oint_S{\bf r}\cdot{\bf P}_{\rm e}\cdot d{\bf S} 
+\sum_\alpha {\bf R}_{\alpha}\!\cdot\!{\bf F}_\alpha\label{e:eossurJ}\\
&\approx& 3N\left[ \frac25\epsilon^0_{\rm F}+n\frac{d\epsilon_{\rm xc}}{dn}\right]_{n=n_0^{\rme}}-N\frac14C\frac{Z_{\rm out}^2}{a} \label{e:PJanak}
\end{eqnarray}
with $\epsilon^0_{\rm F}$ being the Fermi energy of a noninteracting electron gas, if the muffin-tin approximation is not introduced. 
In the derivation of (\ref{e:PJanak}), the electron density $n$ is assumed to be constant $n_0^{\rm e}$ at the surface of the unit cell $\Omega_0$ and 
we used the following relations
\begin{equation}
T_s[n]_{\Omega_0}-T_s[n_0^{\rm e}]_{\Omega_0}
=\frac12\int_{\Omega_0} n({\bf r}){\bf r}\nabla 
U_{\rm eff}({\bf r})d{\bf r}\,, \label{e:Tsall}
\end{equation}
and 
\begin{equation}
T_s[n_{\rm b}]_{\Omega_0}
=\frac12\int_{\Omega_0} n_{\rm b}({\bf r}){\bf r}\nabla 
U_{\rm eff}({\bf r})d{\bf r}\,, \label{e:Tsb}
\end{equation}
for the bound-electron density $n_{\rm b}({\bf r})$. 
Equation~(\ref{e:PJanak}) indicates an essential structure of his pressure formula, 
although simplified.
For the same reason, his formula [his (25)] for the total energy becomes
\begin{equation}
\fl E=-\frac{N}2\int_{\Omega_0} n({\bf r}){\bf r}\nabla 
U_{\rm eff}({\bf r})d{\bf r}+E[n_0^{\rm e}]
-\{\Delta T_{\rm xc}[n({\bf r})]-\Delta T_{\rm xc}[n_0^{\rm e}]\}
-N\frac14C\frac{Z_{\rm out}^2}{a}\,,
\end{equation}
where $\Delta T_{\rm xc}[n({\bf r})]$ in the LDA is given by (\ref{e:dTxc}).\\

\section{Pressure in liquid metals and plasmas}\label{s:ThP}
In Sec.~\ref{s:EPressure}, we have derived pressure formulas for the 
electrons in the fixed nuclei. However, the thermodynamical pressure  of liquid metals and plasmas must be 
determined as a nucleus-electron mixture as a whole; the electron pressure $P_e$ is only a part of the whole pressure $P$, 
although $P_e$ is nearly equal to the total pressure in metals.

In general, the thermodynamic pressure is defined by $P=-\partial F/\partial \Omega|_{N,T}$ using the free-energy of the system: $F=-k_{\rm B}T\ln Z_{N}$ 
with the partition function $Z_{N}={\rm Tr} \exp(-\beta\hat H)$ of a 
nucleus-electron mixture. Here, its Hamiltonian $\hat H\equiv 
\hat H_{\rm ee}+\hat H_{\rm en}+\hat H_{\rm nn}$ consists of the electron and nucleus Hamiltonians 
$\hat H_{\rm ee}$ and $\hat H_{\rm nn}$, in conjunction with 
the electron-nucleus interaction Hamiltonian $\hat H_{\rm en}$. 
When the nuclei can be treated as classical particles 
$[{\bf P}_\alpha,{\bf R}_\alpha]=0$, 
the partition function is written in the form 
(see \cite{Junif} for details)
\begin{equation}
Z_N= \frac{1}{N!h^{3N}}\int\!\! d{\bf R}^N\!\!\int 
d{\bf P}^N\exp[-\beta H_{\rm eff}]
\end{equation}
with
\begin{equation}
H_{\rm eff}
=\sum_\alpha \frac{{\bf P}_\alpha^2}{2M}+\Phi(\{{\bf R}_\alpha\};n)\,,
\end{equation}
\begin{equation}\label{e:efree1}
\Phi(\{{\bf R}_\alpha\};n)\equiv {\cal F}_0+E_{\rm es}[{\hat n}_{\rm I},n]
+{\cal F}_{\rm xc}[{\hat n}_{\rm I},n]
={\cal F}_{\rm e}[\{{\bf R}_\alpha\};n]\,.
\end{equation}
Here,  ${\hat n}_{\rm I}({\bf r})=\sum_\alpha\delta ({\bf r}-{\bf R}_\alpha)$ 
denotes the microscopic nuclear density distribution. Also, it is important 
to realise that $\Phi(\{{\bf R}_\alpha\};n)$ plays a role of a many-body interaction 
among the nuclei, and ${\cal F}_{\rm e}[\{{\bf R}_\alpha\};n]
=\Phi(\{{\bf R}_\alpha\};n)$ represents at the same time the free energy of 
the electrons under the external potential caused by the fixed nuclei at 
the positions $\{{\bf R}_\alpha\}$ in the DF theory 
with reference to ${\cal F}_0$, the free-energy of noninteracting electron gas.

A volume change of the free energy can be calculated by introducing a
new variable $\lambda$ defined by $\Omega(\lambda)=\lambda^3\Omega$ 
in the following way
\begin{equation}
\Omega\frac{\partial F}{\partial \Omega}=\left.\Omega(\lambda)\frac{\partial F(\lambda)}
{\partial \lambda}\frac{\partial \lambda}{\partial \Omega(\lambda)}\right|_{\lambda=1}=
\left.\frac\lambda3\frac{\partial F}{\partial \lambda}\right|_{\lambda=1}\,.
\end{equation}
Here, $F(\lambda)\equiv -k_{\rm B}T\ln Z_{N}(\lambda)$\ is defined by  
\begin{eqnarray}
Z_N(\Omega)&=& \frac{1}{N!h^{3N}}\int_\Omega\!\! d{\bf R}^N\!\!\int 
d{\bf P}^N\exp[-\beta H_{\rm eff}({\bf R},{\bf P})]\\
&=&\frac{\Omega(\lambda)^N}{N!h^{3N}}\int_0^1\!\! d\bar{\bf R}^N\!\!\int 
d{\bf P}^N\exp[-\beta H_{\rm eff}(\eta \bar{\bf R},{\bf P})]\equiv Z_N(\lambda)
\end{eqnarray}
with $\bar {\bf R}={\bf R}/\eta(\lambda)$ and
$\eta^3(\lambda)\equiv\Omega(\lambda)$. Then, the $\lambda$-derivative of $F$  
is easily performed as follows \cite{Hill}
\begin{eqnarray}
{\partial F(\lambda)\over \partial \lambda}
&=&-\frac{Nk_{\rm B}T}{\Omega}\frac{\partial\Omega}{\partial\lambda}\nonumber\\
&&+\frac{\Omega(\lambda)^N}{Z_N(\lambda)}\int_0^1\!\! d\bar{\bf R}^N\!\!\int 
d{\bf P}^N\exp[-\beta H_{\rm eff}(\eta \bar{\bf R},{\bf P})]
\frac{\partial H_{\rm eff}}{\partial \lambda}
\end{eqnarray}
with
\begin{eqnarray}
\left.\frac{\partial H_{\rm eff}}{\partial \lambda}\right|_{\lambda=1}
&=&\left.\frac{\partial \Phi(\eta \bar{\bf R},\Omega(\lambda))}
{\partial \lambda}\right|_{\lambda=1}
=\sum_\alpha {\bf R}_{\alpha}\nabla_\alpha \Phi+3\Omega\frac{\partial \Phi}
{\partial \Omega}\,.
\end{eqnarray}
Finally as a consequence of the above equation, the thermodynamical pressure 
of liquid (solid) metals and plasmas is 
established as a mixture of nuclei and electrons in the form
\begin{eqnarray}
3P\Omega&=&-3\Omega\frac{\partial F}{\partial \Omega}=-\left.\lambda\frac{\partial F}
{\partial \lambda}\right|_{\lambda=1}\\
&=&3Nk_{\rm B}T+\frac1\Xi\int d{\bf R}^N\exp[-\beta \Phi]\left\{ 
-3\Omega\frac{\partial \Phi}{\partial \Omega}+\sum_\alpha 
{\bf R}_{\alpha}\!\cdot\!{\bf F}_\alpha\right\} \label{e:truevir0}\\
&=&3Nk_{\rm B}T+\left< 
-3\Omega\frac{\partial \Phi}{\partial \Omega}\right>
+\left< \sum_\alpha {\bf R}_{\alpha}\!\cdot\!{\bf F}_\alpha\right> 
\label{e:truevir}
\end{eqnarray}
with
\begin{equation}
\Xi\equiv \frac{1}{N!\Lambda^{3N}}\int d{\bf R}^N\exp[-\beta \Phi(\{{\bf R}_\alpha\})]\,,
\end{equation}
and $\Lambda$ being the de Broglie thermal wavelength of the nuclei.
Here, $-\Omega{\partial \Phi}/{\partial \Omega}=P_{\rm e}\Omega$ is 
the electron-pressure in the fixed nuclei, which can be determined 
by (\ref{e:eeossur}) or (\ref{e:eeosvol}) within the framework of the DF theory, 
since $\Phi={\cal F}_{\rm e}[\{{\bf R}_\alpha\}]$ is 
the free energy of electrons in the presence of nuclei fixed at $\{{\bf R}_\alpha\}$. 
At this stage, an important remark should be made concerning (\ref{e:truevir0}). 
It has long been recognised \cite{Janak74,Averill81,NieMar85,ZiescheGN88} that the virial theorem for a liquid (or solid) metal is written 
in the form
\begin{equation}\label{e:Pslater2}
3P_{\rm e}\Omega\equiv \sum_\alpha {\bf R}_{\alpha}\!\cdot\!{\bf F}_\alpha
=2T_\Omega+U_\Omega\,.
\end{equation} 
as was done by Slater \cite{SlaterBK}. However, Eq.~(\ref{e:truevir0}) indicates 
that the electron pressure under the external potential caused by fixed nuclei 
should be given by
\begin{equation}\label{e:PeTKU}
3P_e \Omega=-3\Omega\frac{\partial \Phi}{\partial \Omega} = 2T_\Omega+U_\Omega
-\sum_{\alpha\epsilon\Omega} {\bf R}_{\alpha}\!\cdot\!{\bf F}_\alpha\,,
\end{equation} 
as was shown in (\ref{e:eeosBad}), and that the thermodynamical pressure $P$ of a liquid (solid) metal 
should be defined by (\ref{e:truevir0}) as a nucleus-electron 
mixture, where the nuclear virial term in (\ref{e:PeTKU}) disappears by cancellation.

Thus, the EOS of the nucleus-electron mixture 
is determined by two types of the formulas using 
a surface integral and a volume integral in the final form
\begin{eqnarray}
3P\Omega&=&3Nk_{\rm B}T+\oint_S{\bf r}\cdot\left<\tilde{\bf P}_{\rm K}^{\rm DF}+{\bf P}_{\rm xc}^{\rm DF}\right>\cdot d{\bf S} +\left<\sum_\alpha {\bf R}_{\alpha}\!\cdot\!{\bf F}_\alpha\right> \label{e:ThPsur}\\
&=&3Nk_{\rm B}T+\left<2T_s[n]+E_{\rm es}+
\int {\rm tr}{\bf P}_{\rm xc}^{\rm DF}d{\bf r}\right>\label{e:ThPvol}\,, 
\end{eqnarray}
with combined use of (\ref{e:truevir0}), (\ref{e:eeossur}) and (\ref{e:eeosvol}).
It is worth pointing out that the volume integral form (\ref{e:ThPvol}) of the virial theorem 
contains the nuclear virial contribution, although it does not appear explicitly comparing with the 
surface integral form (\ref{e:ThPsur}). 
In addition, it should be kept in mind here that the term ${\bf P}_{\rm xc}^{\rm DF}(\hat n_{\rm I},n)$ 
depends on the ion-configuration $\hat n_{\rm I}({\bf r})$, 
and therefore it is not that of the jellium model in a strict sense. 
In the jellium model with use of the LDA, the exchange-correlation pressure is given by $P_{\rm xc, \mu\nu}^{\rm DF}=n^2d\epsilon_{\rm xc}/dn\delta_{\mu\nu}$. In this approximation, we obtain the following expressions
\begin{equation} 
\int_\Omega {\rm tr}{\bf P}_{\rm xc}^{\rm DF}d{\bf r}\approx 3\int_\Omega n^2\frac{d\epsilon_{\rm xc}}{dn}d{\bf r}
\end{equation}
\begin{equation}
\oint_S{\bf r}\cdot\left<\tilde{\bf P}_{\rm K}^{\rm DF}+{\bf P}_{\rm xc}^{\rm DF}\right>\cdot d{\bf S}\approx 3\Omega\left[ n\frac25\epsilon^0_{\rm F}+n^2\frac{d\epsilon_{\rm xc}}{dn}\right]_{n=n_0^e}\,,
\end{equation}
where the electron density is assumed to be constant $n_0^e$ on the surface of the volume $\Omega$.

On the other hand, in liquid metals and plasmas where their effective interactions among ions with uniform density $n_0^{\rm I}$ can be described by a pair interaction $v_{\rm II}^{\rm eff}(r)$, the average of the nuclear virial is given by
\begin{equation}
\left<\sum_\alpha {\bf R}_{\alpha}\!\cdot\!{\bf F}_\alpha\right>
=-\frac12Nn_0^{\rm I}\int g_{\rm II}(r){\bf r}\nabla 
v_{\rm II}^{\rm eff}(r)d{\bf r}\,.
\end{equation}
Therefore, the EOS of liquid metals and plasmas is described in this approximation 
by a formula 
\begin{eqnarray}
P&=&n_0^{\rm I}k_{\rm B}T+\frac23\frac{T_\Omega}
\Omega\nonumber\\
&&+\frac16 n_0^{\rm I}\int g_{\rm ee}^0(r)
v_{\rm ee}(r)d{\bf r}-\frac16 (n_0^{\rm I})^2
\int g_{\rm II}(r){\bf r}\nabla v_{\rm II}^{\rm eff}(r)d{\bf r}\label{e:fleos1}\\
&=&n_0^{\rm I}k_{\rm B}T+ \left.n^2\frac{d}{dn}\!\left( 
\frac35\epsilon^0_{\rm F}+\epsilon_{\rm xc}\right)\right|_{n_0^e}
-\frac16 (n_0^{\rm I})^2\int g_{\rm II}(r){\bf r}
\nabla v_{\rm II}^{\rm eff}(r)d{\bf r} \,.\label{e:fleos2}
\end{eqnarray}
Here, $g^0_{ee}(r)$ is a free electron part of the electron-electron radial 
distribution function \cite{JC00} defined by 
\begin{eqnarray}\label{e:S0ee}
\fl 1+n_0^e{\cal F}_Q[g^0_{ee}(r)\!-\!1]\equiv -{1 \over \pi}\int^\infty_{-\infty}\!\!\!\!d\omega
{\hbar \beta \over 1\!-\!\exp(\!-\!\beta\hbar\omega)}
 \Im \frac{\chi_Q^{0}[\omega]}{1\!+\!n_0^e\beta v_{\rm ee}(Q)
[1\!-\!G(Q,\omega)]\chi^{0}_Q[\omega]}\,,
\end{eqnarray}
in terms of the dynamical local-field correction $G(Q,\omega)$ in the electron-ion mixture. 
In the jellium approximation,  we obtain expressions, $G(Q,\omega)\approx G^{\rm jell}(Q,\omega)$ and 
\begin{equation}
\left(\frac{T_\Omega}
\Omega\right)_{\rm jell}=\left.n\left(\frac35\epsilon^0_{\rm F}+ 3n\frac{d\epsilon_{\rm xc}}{dn}-\epsilon_{\rm xc}\right)\right|_{n_0^e}
=\left.n\left(\frac35\epsilon^0_{\rm F}+ 3\mu_{\rm xc}-4\epsilon_{\rm xc}\right)\right|_{n_0^e}
\end{equation}
to be used in (\ref{e:fleos1}) leading to (\ref{e:fleos2}).
At this point, it should be noted that when we approximate $\Phi(\{{\bf R}_\alpha\})$ at earlier stage to be
\begin{equation}\label{e:efree2}
\Phi(\{{\bf R}_\alpha\})\approx U_0(\Omega)
+\frac12\sum_{\alpha\neq \beta}v_{\rm II}^{\rm eff}(|{\bf R}_\alpha-{\bf R}_\beta|)\,,
\end{equation}
we obtain another EOS 
\begin{equation}\label{e:fleos3}
\frac{\beta P}{n_0^{\rm I}}=1+ n_0^{\rm I}\frac{dU_0}{dn_0^{\rm I}}
-\frac16 \beta n_0^{\rm I}\int g_{\rm II}(r)\left[{\bf r}
\nabla -3n_0^{\rm I}\frac{\partial}{\partial n_0^{\rm I}}\right]v_{\rm II}^{\rm eff}(r)d{\bf r}\,,
\end{equation}
which is used as a standard virial formula for a liquid metal \cite{HansenMc}.
The difference between (\ref{e:fleos2}) and (\ref{e:fleos3}) comes from the 
electron-pressure term in (\ref{e:truevir}). 
This term with an approximate expression to $\Phi$ (\ref{e:efree2})
leads to (\ref{e:fleos3}), while this term with use of the exact expression (\ref{e:eeossur}) 
yields a more accurate and simpler pressure-formula (\ref{e:fleos2}).

\section{Summary and discussion}
As a first step, we have derived pressure formulas, (\ref{e:eeossur}) 
and (\ref{e:eeosvol}), for the electrons under the external potential produced by fixed nuclei on the basis of the DF theory; one is a surface integral form and the other, a volume integral form showing the exact virial theorem (\ref{e:eeosBad}). The electron pressure tensor is represented by a sum of 
the kinetic tensor and exchange-correlation pressure tensor, ${\bf P}_{\rm e}=\tilde{\bf P}_{\rm K}^{\rm DF}+{\bf P}_{\rm xc}^{\rm DF}$, in the DF theory.

As a second step, the thermodynamical pressure $P$ is defined by the volume derivative of the free energy for a nucleus-electron mixture 
(liquid metals or plasmas) on the assumption that the nuclei behave as classical particles. This procedure leads to the EOS formula given by (\ref{e:truevir0}), which contains the electron-pressure in the presence of fixed nuclei and the nuclear virial. In final, we obtain pressure formulas, (\ref{e:ThPsur}) and (\ref{e:ThPvol}), for liquid metals and plasmas in the surface integral and  volume integral forms, respectively.
In this respect, it is important to recognise that the exchange-correlation pressure tensor ${\bf P}_{\rm xc}^{\rm DF}(\hat n_{\rm I},n)$ is that of the electron-ion mixture, which is approximated by that of the jellium model in the standard treatment. 

For a liquid metal, when we can approximate a many-body interaction $\Phi(\{{\bf R}_\alpha\};n)$ (\ref{e:efree1}) among the nuclei to be represented by the sum of a binary interaction $v_{\rm II}^{\rm eff}(r)$, Eq.~(\ref{e:ThPsur}) is reduced to a virial pressure formula (\ref{e:fleos1}).
Furthermore, when the electrons in the mixture can be treated by 
the jellium model, (that is, the electron is weakly coupled with the ions in the electron-ion mixture), we obtain a virial pressure formula 
(\ref{e:fleos2}), which is simpler and more accurate than 
the standard virial formula (\ref{e:fleos3}) for a liquid metal \cite{HansenMc}.

Also, our formulation for the pressure have clarified confusion 
and improper expressions found in the previous works, as was addressed in the introduction. The electron pressure in the presence of fixed nuclei should be determined by (\ref{e:PeTKU}) instead of (\ref{e:Pslater2}), as was discussed in Sec.~\ref{s:ThP}. 
Also, a tensor, ${\bf P}_e^{\rm NM}\equiv \tilde{\bf P}_{\rm K}^{\rm DF}+{\bf P}_{\rm xc}^{\rm DF}+{\bf P}^{\rm M}$ with the Maxwell pressure tensor ${\bf P}^{\rm M}$, can be interpreted as a pressure tensor only when we treat a subspace $\Omega$ of the system fulfilling the condition $\oint_S {\bf r}\cdot{\bf P}^{\rm M}\cdot d{\bf S}=0$ on its surface.
Therefore, there is no need to introduce a correction term due to the Maxwell pressure tensor in the Liberman pressure expression \cite{Liberman71}, in spite of the suggestion by Follland \cite{Folland86} and Nielsen and Martin \cite{NieMar85}.

There is a fundamental problem whether we can choose arbitrarily 
any subspace in the system for the virial theorem to be applicable.
Srebrenik and Bader \cite{SrebBader75} have answered to this question by giving the zero-flux surface condition to a subspace; a subspace has a close surface through which the flux of $\nabla n({\bf r})$ is everywhere zero, that is,
\begin{equation}
\nabla n({\bf r})\cdot{\bf e}({\bf r})=0\,,
\end{equation}
for any point ${\bf r}$ in a close surface with a normal vector ${\bf e}({\bf r})$.
In a local virial theorem constructed on this condition, there is no 
ambiguity in the definition of kinetic energy $T_s$ and the kinetic tensor ${\bf P}_{\rm K}^{\rm DF}$, as will be discussed in Appendix.
\newpage

\section*{Appendix. Proof of (\ref{e:Ts2})} 
\appendix
\setcounter{section}{1}
In the DF theory, the wave function is determined by 
the Schr\"odinger equation with a self-consistent external potential 
$U_{\rm eff}({\bf r})$
\begin{equation}
\left[-\frac{\hbar^2}{2m}\nabla^2+U_{\rm eff}({\bf r}) -\epsilon_i 
\right]\varphi_i({\bf 
r})=0 \label{e:we2}\,.
\end{equation}
By multiplying (\ref{e:we2}) by $\varphi_i^*{\bf r}\nabla$, there results
\begin{eqnarray}
\fl \varphi_i^*{\bf r}\nabla\left[-\frac{\hbar^2}{2m}\nabla^2+U_{\rm eff}
({\bf r}) -\epsilon_i \right]\varphi_i({\bf r})
&=&\frac{-\hbar^2}{2m}\varphi_i^*{\bf r}\nabla \nabla^2\varphi_i+
\varphi_i^*{\bf r}\nabla U_{\rm eff}\varphi_i
\nonumber\\
&&+[U_{\rm eff} -\epsilon_i]\varphi_i^*{\bf r}\nabla\varphi_i=0\,.
\end{eqnarray}
Therefore, from the above equation \cite{Slater72}, we obtain 
\begin{eqnarray}
\varphi_i^*\varphi_i{\bf r}\nabla U_{\rm eff}
&=&\frac{\hbar^2}{2m}[\varphi_i^*{\bf r}\nabla \nabla^2\varphi_i-
(\nabla^2 \varphi_i^*)({\rm r}\nabla \varphi_i)]
\nonumber\\
&=&\frac{-\hbar^2}{2m}\left[2\varphi_i^* \nabla^2\varphi_i-
\nabla\{ (\varphi_i^*)^2\nabla ({\bf r}\varphi_i/\varphi_i^*)\}\right]\,.
\end{eqnarray}
This equation is rewritten in the form
\begin{eqnarray}
2\varphi_i^* \frac{-\hbar^2}{2m}\nabla^2\varphi_i
=\varphi_i^*\varphi_i{\bf r}\nabla U_{\rm eff}+A
\end{eqnarray}
with
\begin{eqnarray}
\fl A&=&\frac{-\hbar^2}{2m}
\nabla\cdot\{\varphi_i^*\nabla[{\bf r}\nabla\varphi_i]
-({\bf r}\nabla\varphi_i)(\nabla\varphi_i^*)\}\\
\fl &=&\frac{-\hbar^2}{4m}
\nabla\cdot\{\varphi_i^*\nabla[{\bf r}\nabla\varphi_i]
-({\bf r}\nabla\varphi_i)(\nabla\varphi_i^*)+{\rm c.c.}\}\\
\hspace{-2cm}&=&\frac{-\hbar^2}{4m}\nabla\cdot\{\nabla (\varphi_i^*\varphi_i)+
({\bf r}\nabla)\nabla (\varphi_i^*\varphi_i)
-2[({\bf r}\nabla\varphi_i)(\nabla\varphi_i^*)+({\bf r}\nabla\varphi_i^*)(\nabla\varphi_i)]\}\\
\fl &=&\frac{-\hbar^2}{4m}\nabla\cdot\{\nabla (\varphi_i^*\varphi_i)
+\varphi_i^*({\bf r}\nabla)\nabla \varphi_i+\varphi_i({\bf r}\nabla)\nabla 
\varphi_i^*\nonumber\\
\fl &&\quad\quad\quad\quad\qquad\qquad -({\bf r}\nabla\varphi_i)(\nabla\varphi_i^*)-({\bf r}\nabla\varphi_i^*)(\nabla\varphi_i)\}\,,
\end{eqnarray}
which leads to (\ref{e:Ts2}) with (\ref{e:sur1})-(\ref{e:sur3}).

On the other hand, (\ref{e:Ts2}) can be proved in a direct way 
by following Pauli \cite{Pauli}. 
Pauli have proved that the probability current
\begin{equation}
{\bf i}\equiv \frac12\left[(\varphi{\frac{\bf \hat p}{m}}\varphi^*)
+(\varphi{\frac{\bf\hat p}{m}}\varphi^*)^*\right]
\end{equation}
for a wave function of a single electron moving 
under the external potential $U_{\rm eff}({\bf r})$
obeys the following continuity equation
\begin{equation}
m\frac\partial{\partial t}{\bf i}=-\nabla \cdot {\bf S}
-(\varphi\varphi^*)\nabla U_{\rm eff}({\bf r})
\end{equation}
with a tensor defined by
\begin{equation}
S_{\mu\nu}(\varphi)\equiv \frac{-\hbar^2}{4m}\left[
\varphi\nabla_\mu\nabla_\nu\varphi^*+\varphi^*\nabla_\mu\nabla_\nu\varphi
-\nabla_\mu\varphi\nabla_\nu\varphi^*-\nabla_\mu\varphi^*\nabla_\nu\varphi \right]\,.
\end{equation}
This provides the following equation for the static case
\begin{equation}\label{e:ProbCont}
m{\bf r}\cdot\frac\partial{\partial t}{\bf i}=-{\bf r}\cdot\nabla \cdot {\bf S}
-(\varphi\varphi^*){\bf r}\nabla U_{\rm eff}({\bf r})=0\,.
\end{equation}
When we interpret ${\bf P}_{\rm K}^{\rm DF}$ defined by
\begin{equation}
{\bf P}_{\rm K}^{\rm DF}\equiv \sum_i f(\epsilon_i){\bf S}(\varphi_i)
\end{equation}
as a kinetic tensor in the DF theory, (\ref{e:ProbCont}
) is changed into the form
\begin{eqnarray}\label{e:virPauli}
-\int_\Omega{\bf r}\cdot\nabla \cdot {\bf P}_{\rm K}^{\rm DF}d{\bf r}
&=&\int_\Omega n({\bf r})\nabla U_{\rm eff}({\bf r})d{\bf r}\nonumber\\
&=&\int_\Omega {\rm tr}{\bf P}_{\rm K}^{\rm DF}d{\bf r}
-\oint_S{\bf r}\cdot{\bf P}_{\rm K}^{\rm DF}\cdot d{\bf S}\,, 
\end{eqnarray}
with use of a tensor relation (\ref{e:tensor}). 
By noting the relation
\begin{eqnarray}
{\rm tr}{\bf S}&=&\sum_\mu S_{\mu\mu}=\frac{\hbar^2}{4m}\left[ 
-\varphi^*\nabla^2\varphi-\varphi\nabla^2\varphi^*
+2\nabla\varphi^*\nabla\varphi \right]\\
&=&\left[(\varphi^*\frac{-\hbar^2}{2m}\nabla^2\varphi)
+(\varphi^*\frac{-\hbar^2}{2m}\nabla^2\varphi)^*\right]
+\frac{\hbar^2}{4m}\nabla^2(\varphi^*\varphi)\\
&=&2\frac{1}{2m}\left|\frac{\hbar}i \nabla\varphi\right|^2-\frac{\hbar^2}{4m}\nabla^2(\varphi^*\varphi)\\
&=& \frac12\left[(\varphi^*\frac{-\hbar^2}{2m}\nabla^2\varphi)
+(\varphi^*\frac{-\hbar^2}{2m}\nabla^2\varphi)^*\right]
+\frac{1}{2m}\left|\frac{\hbar}i \nabla\varphi\right|^2\,,
\end{eqnarray}
we can represent the trace of ${\bf P}_{\rm K}^{\rm DF}$ in the form
\begin{equation}\label{e:trKEpauli}
\int_\Omega{\rm tr}{\bf P}_{\rm K}^{\rm DF} d{\bf r}
=2T_s[n]_\Omega+\frac{\hbar^2}{4m}\oint_S \nabla n({\bf r})\cdot d{\bf S}\,.
\end{equation}
From (\ref{e:virPauli}) and (\ref{e:trKEpauli}), we obtain in final
\begin{eqnarray}\label{e:TsS}
2T_s[n]_\Omega
&=&\int_\Omega n({\bf r})\nabla U_{\rm eff}({\bf r})d{\bf r}\nonumber\\
 &&+\oint_S{\bf r}\cdot{\bf P}_{\rm K}^{\rm DF}\cdot d{\bf S}
 -\frac{\hbar^2}{4m}\oint_S \nabla n({\bf r})\cdot d{\bf S}\,,
\end{eqnarray}
which is identical with (\ref{e:Ts2}) with (\ref{e:sur3}).
In this respect, it is interesting to note that 
we have arbitrariness to define the kinetic-energy functional in another form
\begin{eqnarray}
2T_s[n]_\Omega&\equiv& \int_\Omega{\rm tr} {\bf P}_{\rm K}^{\rm DF}d{\bf r}\\
 &=&\sum_i f(\epsilon_i)\int_\Omega\left[(\varphi_i^*\frac{-\hbar^2}{2m}\nabla^2\varphi_i)
+(\varphi_i^*\frac{-\hbar^2}{2m}\nabla^2\varphi_i)^*\right]d{\bf r}\nonumber\\
 &&+\frac{\hbar^2}{4m}\int_\Omega\nabla^2n({\bf r})d{\bf r}\,,
\end{eqnarray}
involving the last term of (\ref{e:TsS}) here.

\end{document}